\documentclass[english]{extarticle}
\usepackage[T1]{fontenc}
\usepackage[latin9]{inputenc}
\usepackage{geometry}
\usepackage{float}
\usepackage{graphicx}
\usepackage[numbers]{natbib}

\makeatletter
\@ifundefined{date}{}{\date{}}
\makeatother

\usepackage{babel}
\begin{document}
\title{Towards ``Weak'' Information Theory: \\Weak-Joint Typicality Decoding
Using Support Vector Machines May Lead to Improved Error Exponents\thanks{This project was commenced in early 2020. It has been partly presented
in initial form in the first author's \emph{Anabhayin} series of self-published
books.}}
\author{Aman Chawla\\Department of Computer Science and Engineering\\Dayananda
Sagar University\\and\\Salvatore Domenic Morgera\\Department of
Electrical Engineering\\University of South Florida}
\maketitle
\begin{abstract}
In this paper, the authors report a way to use concepts from statistical
learning to gain an advantage in terms of error exponents while communicating
over a discrete memoryless channel. The study utilizes the simulation
capability of the scientific computing package MATLAB to show that
the proposed decoding method performs better than the traditional
method of joint typicality decoding. The advantage is secured by modifying
the traditional specification of what constitutes a decoding error.
This is justified by the paradigm, also used in the program of `utilizing'
noisy feedback, that one ought not to declare a condition as an error
if some further processing can extract useful information from it.
\end{abstract}

\part{Theoretical Foundations}

Noisy feedback has been found useful in recent information theory
literature\footnote{See, for example, the papers of Prof. Anant Sahai (UC Berkeley) and
Prof. Sekhar Tatikonda (Yale).}. In the same spirit, we investigate the potential use of discarded
``erroneous'' conditions in this paper. In this part we provide some
theoretical underpinnings of our computer investigations. The investigations
relate to the concept of error exponents. These functions capture
the performance of a communication system in a subtle and important
way. Along with channel capacity, the error exponent is an important
metric of communication system performance. 

\section{Brief Outline}

Consider Theorem 7.7.1 of Cover and Thomas, page 200 \citep{Cover2005}.
In the proof of this theorem, the authors lay out a few points. The
sixth point details how \emph{joint typicality decoding} is to be
used in the proof. Motivated by `weak values' in quantum measurement
theory and `strong learning' and `weak learning' in the book on machine
learning by Kearns and Vazirani \citep{kearns1994introduction}, we
propose to `weaken' this decoding step. Specifically, in weak-joint
typicality decoding, the receiver declares that the index $\hat{W}$
was sent if the following conditions are satisfied:
\begin{enumerate}
\item $(X^{n}(\hat{W}),Y^{n})$ are jointly typical 
\item There may or may not be other indices $W'\sim=\hat{W}$ such that
$(X^{n}(W'),Y^{n})$ belongs to the jointly typical set
\end{enumerate}
This is a sort of fuzzy decoding set and gives us the idea that neural
networks could `learn' in this setting. If no such $\hat{W}$ exists,
satisfying condition 1. above, then an error is declared and the receiver
outputs a dummy index such as zero in this case.

Suppose we decode to $\hat{W}$ since $(X^{n}(\hat{W}),Y^{n})$ is
a jointly typical pair. We also find that $\hat{W}_{k}$, $k=1,2,...,L$
are such that $(X^{n}(\hat{W}_{k}),Y^{n})$ are jointly typical. So
we have $L+1$ decodings, potentially. We can look for clusters in
this data via unsupervised learning, and possibly decode to a point
chosen at random from the largest cluster\footnote{See k-means clustering from Haykin's book, \emph{Neural Networks and
Learning Machines}, page 242, section 5.5 \citep{haykin2010neural}.}. Following this set up, we can finally evaluate the probability of
error so obtained.

In the next section we will provide the details of the transmission
scheme which will be computerized in the next part. 

\section{Details of the Transmission Scheme}

Consider a communication system, the transmission scheme over which
consists of $M$ messages, belonging to the set $\{W_{1},W_{2},W_{3},...,W_{M}\}$.
The encoder maps these $M$ messages into input sequences for the
channel. These input sequences, or codewords, belong to the set $\{X^{n}(W_{i})\}_{i=1}^{M}$.
Each member of this latter set is of length $n$ and there are $M$
members. One of these codewords is transmitted per cycle. Due to channel
noise, it is received at the receiver in the form $Y^{n}$. 

Now suppose that the transmitted sequence is unknown, and we only
have access to $Y^{n}$. The question is how do we decode the message
hidden in $Y^{n}$. In particular, suppose that under joint typicality
decoding, two distinct transmitted sequences $X^{n}(W_{1})$ and $X^{n}(W_{6})$
are jointly typical with $Y^{n}$. This raises a quandry for the decoder.
Suppose there is a `difference function' on sequence space, $d(\cdot,\cdot)$
which yields a new sequence which is the difference between the two
input sequences. For example, it might be based on binary Hamming
distance. Let $Z_{i}^{n}=d(X^{n}(W_{i}),Y^{n})$ for $i=1,6$. Whereas
the codeword is deterministic\footnote{However, if the codebook is also chosen randomly as in the proof of
Theorem 7.7.1 \citep{Cover2005}, then the codeword is no longer deterministic.
Furthermore, we may draw the message also at random using a uniform
distribution over the message set.}, the received sequence is random due to the noise and so the formed
$Z$-sequences are also random. We can then use unsupervised learning
to form $k$ clusters out of these $Z$-sequences. We then search
for and find the largest among the $k$ clusters. Next we find its
mean and denote it by $m$. Finally, we decode to that $Z$-sequence
which is closest to $m$.

In the next part, we will perform MATLAB simulations of communication
over a discrete channel and perform decoding as outlined above, as
well as the more standard joint typicality decoding.

\part{Computer Experimentation}

In this part, just as laid out earlier in this paper, we relax the
error-event definition in joint typicality decoding over a noisy channel
and invoke Support Vector Machines (SVMs) to help us decode in the
presence of several \textquoteleft matches\textquoteright{} between
the received word and codewords from the codebook\footnote{Thanks are due to Prof. Rajesh, CSE, DSU for suggesting the use of
SVMs instead of k-means clustering due to the vector nature of the
codewords.}. We can also extend this to source coding and rate distortion theory
in future work. 

\section{Study of Error Exponents}

\begin{figure}[H]
\caption{Joint typicality decoding error exponent versus block length\label{fig:Joint-Typicality-Decoding}.}

\begin{raggedright}
\includegraphics[width=6in]{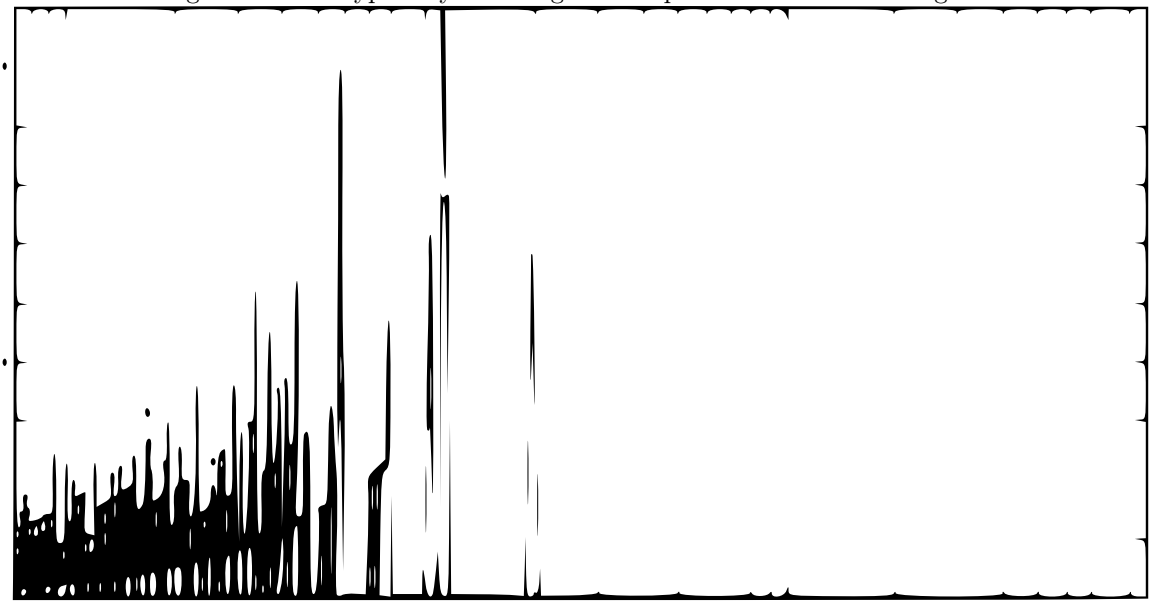}
\par\end{raggedright}
\end{figure}

In Figure \ref{fig:Joint-Typicality-Decoding}, we plot the formula 

\begin{equation}
-\frac{ln(Pe)}{blocklength},\label{eq:errorexponentdefinition}
\end{equation}
for various blocklengths, for the discrete memoryless binary symmetric
channel BSC(0.05). This is done for the case of joint typicality decoding.
The x-axis is the block length simulated and the y-axis is the error
exponent.

\begin{figure}[H]
\caption{Weak-joint typicality decoding error exponent versus block length. The y-axis peak value is nearly double that of the y-axis peak value in Figure \ref{fig:Joint-Typicality-Decoding}\label{fig:Weak-Decoding-Error}.}

\raggedright{}\includegraphics[width=6in]{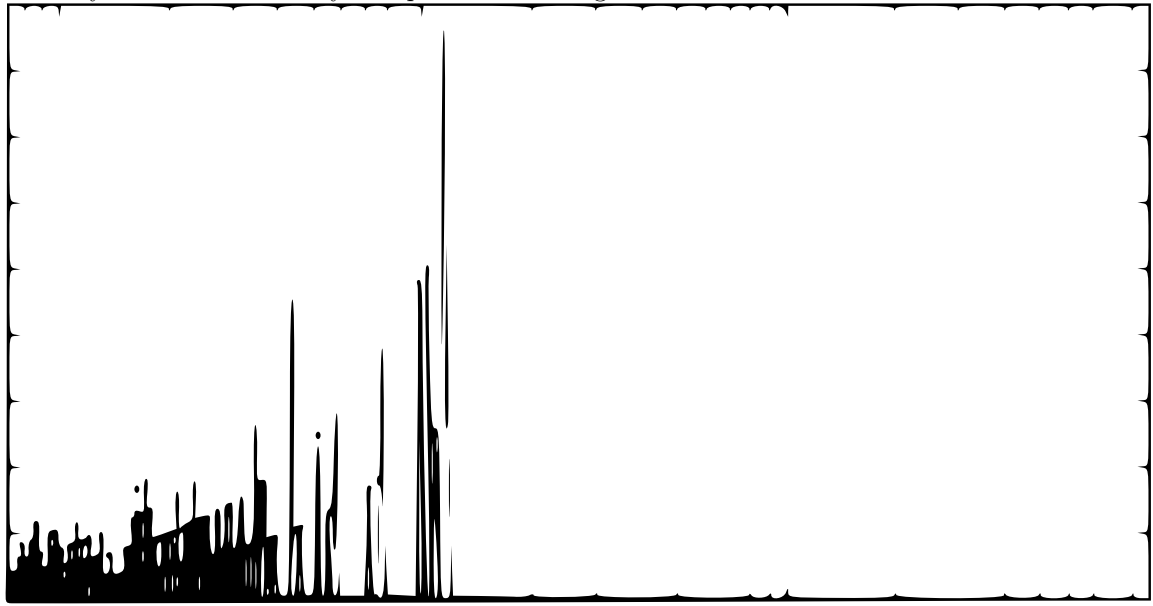}
\end{figure}

Next, in Figure \ref{fig:Weak-Decoding-Error}, we plot the same formula
{[}\ref{eq:errorexponentdefinition}{]}, but for weak-joint typicality
decoding, and for the same channel. We can see that the exponent is
higher for certain rates\footnote{Recall that the rate is nothing but the number of bits carried by
a codeword divided by the length of the codeword.}. The x-axis is the block length simulated and the y-axis is the error
exponent.

\section{Results}

\begin{figure}[H]
\caption{The advantage of weak-joint typicality decoding is in terms of error
exponents.\label{fig:The-advantage-of}}

\raggedright{}\includegraphics[width=6in]{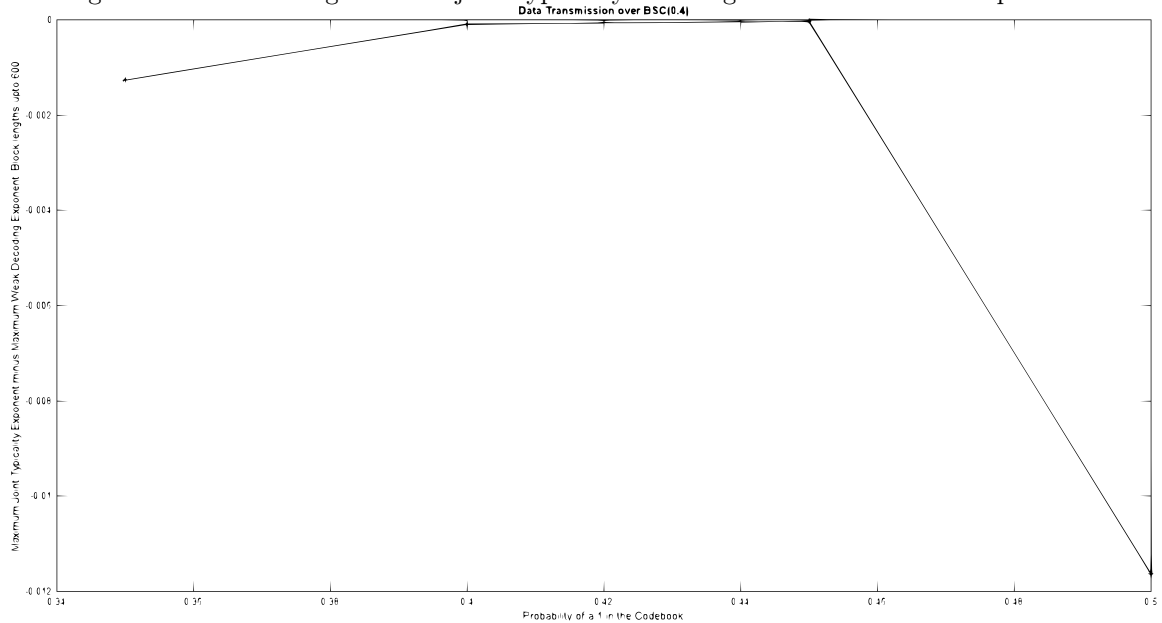}
\end{figure}

In this section we present a figure that summarizes the performance
of the new decoding method. As Figure \ref{fig:The-advantage-of}
shows, there is a definite advantage in using weak-joint typicality
decoding over regular joint typicality decoding. The x-axis is the
probability of a `1' symbol in the codebook and the y-axis is the
difference between the (maximum, simulation-obtained, value of the)
two types of decoding-exponents, with a block length of upto 600 symbols
considered for codeword length. The graph always remains below zero,
demonstrating the greater magnitude of the (maximum) weak-joint typicality
decoding exponent. The discrete channel used was the BSC(0.4). 

\section{Discussion}

While the differences are not glaring with and without weak-joint
typicality decoding, we have obtained a modification of Shannon\textquoteright s
strategy that performs better in terms of error exponents, under certain
conditions. It remains to do an analytical study. Additionally, it
was suggested recently that our strategy might be related to list
decoding \citep{forney1968exponential}. However, a preliminary reading
of \citep{gallager1968information} indicates that there might be
some differences between these two types of decoding. 

To summarize, in this paper we have discussed a potentially new way
of using machine learning to enhance communication system performance
as captured by error exponents. There are limitations to our work,
both in theory and simulation. Theory can be improved by considering
other machine learning algorithms and the simulations can be more
exhaustive. However, this is a \emph{preliminary study} of new frontiers
in information theory and statistical learning. It has several practical
implications for those implementing data communication systems in
the twenty-first century.

\section*{Acknowledgments}

The first author would like to thank Prof. Baris Nakiboglu of METU,
Turkey for his critical comments and pointing out the relation with
list decoding.

\bibliographystyle{unsrtnat}
\bibliography{mylib2}

\end{document}